\documentclass[onecolumn,showpacs,preprintnumbers,amsmath,amssymb]{revtex4}
\usepackage{graphicx}% Include figure files
\usepackage{dcolumn}% Align table columns on decimal point
\usepackage{bm}% bold math
\begin{document}

%\preprint{APS/123-QED}

\title{Electromagnetic properties of the $d^*(2380)$ hexaquark}% Force line breaks with \\
\author{M. Bashkanov}
\email{mikhail.bashkanov@york.ac.uk}
 \affiliation{Department of Physics, University of York, Heslington, York, Y010 5DD, UK}%Lines break automatically or can be forced with \\
\author{D.P. Watts}%
\affiliation{Department of Physics, University of York, Heslington, York, Y010 5DD, UK}%Lines break automatically or can be forced with \\
\author{A. Pastore}%
\affiliation{Department of Physics, University of York, Heslington, York, Y010 5DD, UK}%Lines break automatically or can be forced with \\


\date{\today}% It is always \today, today,

\begin{abstract}
Experiments with intense photon and electron beams have the potential to provide access to nontrivial properties of the recently discovered $d^*(2380)$ hexaquark including its size, structure, magnetic moment, quadrupole and octupole deformations. In this paper we investigate the sensitivity of ongoing and planned experiments to various properties of the $d^*(2380)$, employing models based on both constituent quark and pion cloud frameworks. Our calculations indicate that for photoinduced reactions on the deuteron the $d^*(2380)$ is predominantly produced from the $D-wave$ component of the deuteron. We confirm earlier findings that the intrinsic quadrupole deformation of the $d^*(2380)$ should be small. We also demonstrate an ability to extract the $d^*(2380)$ magnetic moment and put constrains on the $d^*(2380)$ $M3/E2$ ratio.
\end{abstract}

\maketitle

%\tableofcontents

\section{\label{sec:Intro} Introduction}

The properties of the $d^*(2380)$ have been established quite rigorously in recent years following its observation in proton-neutron scattering and pionic fusion reactions~\cite{mb,MB,MBC,TS1,TS2,MBA,MBE1,MBE2,BCS}. It has a mass of $M_{d^*}=2380$~MeV, vacuum width $\Gamma = 70$~MeV and quantum numbers $I(J^P)=0(3^+)$. It therefore provides a new bosonic, isoscalar configuration in the light-quark sector. The internal structure of the $d^*(2380)$ is quite complicated and to some extent resembles a deuteron in which the nucleons are substituted with $\Delta$'s, where $\Delta$ refers the lowest lying excited state of the nucleon. The wavefunction in such a case is given by $|\Psi_{d^*}> =|6q> +|\Delta\Delta_{S-wave}>+|\Delta\Delta_{D-wave}>$. Each of the components in the wavefunction has its own spatial  extension~\cite{DBT9} with the $|6q>$ configuration the most compact one $R_{d^*}(|6q>)\sim 0.5$~fm ($R_{d^*}(|\Delta\Delta_{S-wave}>)\sim 0.8$~fm and  $R_{d^*}(|\Delta\Delta_{D-wave}>)\sim 1.4$~fm). The $|6q>$ compact configuration of the $d^*(2380)$ is predicted to be dominant ($\sim 69\%$) while the more extended $D-wave$ "molecular" component is of order $~2\%$.

In a very recent work the possibility of $d^*(2380)$ formation within neutron stars was explored in relativistic mean field calculations where the $d^*(2380)$ was introduced as a diffuse non-interacting and non-condensing gas alongside the standard (nucleonic and leptonic) constituents of neutron stars. By solving Tolman-Oppenheimer-Volkoff (TOV) equations using the resulting Equation of State, it was observed a significant $d^*$ formation  up to 20\% of the matter in the centre of heavy stars predicted to dwell as  $d^*$. The resulting mass-radius predictions for neutron stars with this $d^*$ degree of freedom is currently one of the few that can simultaneously give agreement with both the mass-radius constraint of the recent merger event observed by LIGO~\cite{Abbott}, while giving agreement with the maximum observed (and inferred from GW data) neutron star mass of $\sim 2.17 M_\odot$~\cite{Metzger}.

The influence of the $d^*(2380)$ electromagnetic properties to neutron stars, like contribution of the $d^*(2380)$ magnetic moment to formation of the neutron star magnetic field makes the question of the $d^*(2380)$ size and structure important to astrophysics as well as nuclear physics. It was recently shown that the $d^*(2380)$ can be produced copiously using photon beams of energy $E_{\gamma}\approx 570 MeV$ from a deuteron target~\cite{JapD,BasD,mbMainz}. Due to its high spin, $J^P_{d^*}=3^+$, the $d^*(2380)$ requires the contribution of higher multipoles ($E2, M3$ or $E4$) to be photoproduced from the deuteron ($I(J^P)=0(1^+)$). The ratios of the strengths of different multipoles in EM transitions are sensitive to the shape and magnetic moments, as already evidenced for the $\Delta^+$, where it was possible to extract the intrinsic quadrupole deformation of the $\Delta$ resonance from ratios of the $E2$ and $M1$ multipoles~\cite{Buch1,Buch2}. In this work we will employ such methodologies to investigate the sensitivities can be achieved for the new $d^*(2380)$. Our calculations are based on geometrical arguments only without any inter-particle interaction. The aim of the current article is to provide rough estimates on some important properties of $d^*$ planned to be measured and thus to motivate more refined theoretical models.

The paper is structured as follows. In section 2 we outline the theoretical framework used to describe the N$\Delta$ transition in the pion cloud model and in section 3 adapt this to the case of the d$^{*}$. Sections 4-7 describe the extraction of the $d^*$ electric quadrupole and magnetic octupole moments as well as transition electromagnetic moments.

\section{The $\Delta$ in a pion-cloud model}

In this section, we discuss the earlier work regarding the $\Delta$ with a view to extending this to the $d^*$ in section 3. It was shown in Ref.~\cite{Buch1} that the wave function of $\Delta$ resonance can be considered as a two component sum $|\Delta> = \alpha'|\Delta'>+\beta'|N\pi>$, where $\Delta'$ is a true compact 3-quark configuration and $N\pi$ is the component deriving from a nucleon plus pion cloud. The coefficient $\alpha'^2$ gives the probability to find the $\Delta$ in a three-quark state, while $\beta'^2$ correspond to the probability for the $\Delta$ to be in a pion-cloud mode.  The coefficients are normalised to unity $\alpha'^2+\beta'^2=1$. To agree with existing experimental data the coefficent $\beta'$ is determined to be -0.52~\cite{Buch1}. The quadrupole moment of the $\Delta$ resonance can be calculated in this model as $<\Psi_{\Delta}^*|\hat{Q}_{\pi}|\Psi_{\Delta}>$ where $\hat{Q}_{\pi}$ is a quadrupole operator of the form:

\begin{eqnarray}
\hat{Q}_{\pi}=e_{\pi} \sqrt{\frac{16\pi}{5}}r_{\pi}^2 Y_0^2(\hat{r}_\pi),
\end{eqnarray}
in which $e_\pi$ is the pion charge operator divided by the unit charge $e$, and $r_\pi$ is the distance between the center of the quark core and the pion. The wave function for the $\Delta^+$ can be written in the form

\begin{eqnarray}
|\Delta^+\uparrow> = \alpha'|\Delta^{+'}\uparrow>+\beta'\frac{1}{3}( 2|p'\uparrow\pi^0Y^1_0>+\sqrt{2}|p'\downarrow\pi^0Y^1_1> +\sqrt{2}|n'\uparrow\pi^+Y^1_0>+|n'\downarrow\pi^+Y^1_1> )
\end{eqnarray}
All coefficients in equation 2 are the simple spin-isospin Clebsch-Gordan couplings. Since both neutral pions (zero charge) and the spherically symmetric quark core do not contribute to the quadrupole deformation  one can simplify the wave function by omitting the irrelevant terms.

\begin{eqnarray}
|\Delta^+\uparrow> = \beta'\frac{1}{3}\left( \sqrt{2}|n'\uparrow\pi^+Y^1_0>+|n'\downarrow\pi^+Y^1_1>\right)
\end{eqnarray}
From this the spectroscopic quadrupole moment of the $\Delta$ can be derived: 

\begin{eqnarray}
Q_{\Delta^+}=-\frac{2}{15}\beta'^2r_\pi^2
\end{eqnarray}
The proton to $\Delta$ transition quadrupole moment can be calculated in a similar manner $<\Psi_{p}^*|\hat{Q}_{\pi}|\Psi_{\Delta}>$, giving a  similar result.

\begin{eqnarray}
Q_{\Delta^+}=Q_{p\rightarrow \Delta^+}
\end{eqnarray}
Here the probability of proton to be in the pion-cloud mode is assumed to be $\beta=0.26$ and $r_\pi=1.77 fm$, determined from the experimental data~\cite{Buch1}.

\section{The $d^*(2380)$ in a pion cloud model}

One can apply the same formalism to calculate the $d^*(2380)$ quadrupole deformation under pure geometrical approach. For the case of the $d^*$ we assume a "Deltaron" structure consisting of two $\Delta$s with wavefuction as outlined in section 2. In this non-interacting case the wave function of the $d^*(2380)$ can be written as

\begin{eqnarray}
|d^*> = |\Delta\Delta> = (\alpha'|\Delta'>+\beta'|N'\pi>)\cdot(\alpha'|\Delta'>+\beta'|N'\pi>)
\end{eqnarray}
creating three major structures in the wave function:
\begin{eqnarray}
|d^*> = A+B+C \nonumber \\
\end{eqnarray}
where:
\begin{eqnarray}
A = \alpha'^2|\Delta'\Delta'> \nonumber \\
B = \alpha'\beta'|\Delta'N'\pi> \nonumber \\
C = \beta'^2|N'N'\pi\pi>
\end{eqnarray}
For the subsequent analysis we will only consider configurations with two pions in iospin $I=0$ state and two nucleons in $I=0$ state, the dominanat configurations which encapsulate 80\% of the available parameter space~\cite{BCS}~\footnote{The case with two pions in $I=1$ state and two nucleons in $I=1$, like $pp\pi^-\pi^0$ and $nn\pi^+\pi^0$ are very interesting since iospin selection rules also imply $l_{\pi\pi}=1$ and $l_{NN}=1$}. 
It is interesting to note that this primitive model gives rise to all the major $d^*$ structures proposed to date. The first term, $A$, can be related to a six quark configuration of the $d^*$. However, it should be noted that this configuration is only to one out of 5 possible $6q$ configurations proposed~\cite{Wang6q,BBC}. The second term resembles a "pion assisted" dibaryon configuration as proposed by Gal {\em et. al.}~\cite{Gal1}. The third term can be further decomposed into two cases: $C1$) two pions in a relative $S-wave$, pion pair in a relative $D-wave$ to the $S=1$ nuclear core and $C2$) two pions in a relative $D-wave$, pion pair in a relative $S-wave$ to the $S=1$ nuclear core. The $C1$ case resembles the $\sigma$-cloud model of the $d^*$ by Kukulin {\em et. al.} ~\cite{Kuk1}. The $C2$ term is analagous to the $D-wave$ $\Delta\Delta$ configuration~\cite{BCS1}.

To calculate the quadrupole deformation of the $d^*$ we will use the same ansatz as for the $\Delta$: $<\Psi_{d^*}^*|\hat{Q}_{\pi}|\Psi_{d^*}>$. Considering the contributions to the $d^*$ wavefunction (equations 7 and 8) The $A$~term is spherically symmetric, hence does not give a contribution to the quadrupole moment. The $B$~term represents the quadrupole moment of a single $\Delta^+$ multiplied by $\alpha'^2$ factor:
\begin{eqnarray}
<B|\hat{Q}_{\pi}|B> = \alpha'^2 \cdot Q_{\Delta^+}=(1-\beta'^2)\cdot Q_{\Delta^+}\approx 0.73 Q_{\Delta^+}
\end{eqnarray}
It can be seen that the $N\Delta\pi$ term is responsible for the non sphericity of the $d^*$, producing a small oblated shape similar in character to that of a single $\Delta$. Recent experiments have set limits on a possible $d^*\rightarrow NN\pi$ decay branch and indicate the $N\Delta\pi$ term in the $d^*$ wave function is likely to be very small, if it exists at all~\cite{TS17}.
The resulting suppression of the $B$-term would further reduce the $d^*$ quadrupole deformation.

The $C1$-term ($D-wave$ $\sigma$ cloud) would not contribute to the quadrupole moment due to the zero net charge of the pion pair. To calculate the $C2$ term we adopt a relative coordinate system in which $\hat{r}_\pi$ is substituted by $\hat{r}_{\pi-\pi}$ - the separation between the two pions. The $r_\pi$ is related to ${r}_{\pi-\pi}$ by  $r_{\pi-\pi}=\sqrt{2}r_{\pi}$ - the relative distance between the pions instead of the pion-core distance. Similarly $e_\pi$ would be transformed to 2: ($e_{\pi^+}-e_{\pi^-}$). To simplify the integration we employ the spherical harmonics addition theorem,  decomposing $Y_0^2(\hat{r}_{\pi-\pi})$ into the product $Y^2_m(\hat{r}_{\pi_1})Y^2_{-m}(-\hat{r}_{\pi_2})$.
\begin{eqnarray}
Y_0^2(\hat{r}_{\pi-\pi})=\sqrt{\frac{5}{16\pi}}\sum_{m=-2}^{2} (-1)^mY^2_m(\hat{r}_{\pi_1})Y^2_{-m}(-\hat{r}_{\pi_2})
\end{eqnarray}
The few non-zero elements in this calculations are shown together with their resulting weights below
\begin{eqnarray}
a) <Y^1_{1}Y^1_{-1}|Y^2_0|Y^1_{1}Y^1_{-1}>: \frac{-1}{16\sqrt{5\pi^3}} \nonumber \\
b) <Y^1_{0}Y^1_{0}|Y^2_0|Y^1_{0}Y^1_{0}>: \frac{-1}{4\sqrt{5\pi^3}} \nonumber  \\
c)  <Y^1_{1}Y^1_{0}|Y^2_0|Y^1_{1}Y^1_{0}>: \frac{1}{8\sqrt{5\pi^3}} \nonumber  \\
d) <Y^1_{1}Y^1_{1}|Y^2_0|Y^1_{1}Y^1_{1}>: \frac{1}{16\sqrt{5\pi^3}} \nonumber  \\
\end{eqnarray}
Summing these contributions gives:
\begin{eqnarray}
<C|\hat{Q}_{\pi}|C> = -0.52 Q_{\Delta^+}
\end{eqnarray}
The double-pion cloud produces a prolate quadrupole deformation, similar to that observed for the case of the nucleon. Note that the $B$ and $C$ terms have opposite signs.
Combining all terms and taking $Q_{\Delta^+}= −0.043 efm^2$ from Ref.~\cite{Pena1}~\footnote{We used $Q_{\Delta^+}= - 0.043 efm^2$ from Ref.~\cite{Pena1} as it is closer to the experimentally determined value, instead of $Q_{\Delta^+}= - 0.113 efm^2$ from Ref.~\cite{Buch2}, calculated within pion cloud formalism.} one gets $Q_{d^*}\approx 0.21Q_{\Delta^+}= 0.009 efm^2$ with a magnitude mainly arising from the $B-C$ term cancellations. It should be noted that as discussed in section 2, experimental decay studies suggest a small contribution from the B-term (N$\Delta \pi$) to the $d^*$ wavefunction.  With this assumption the quadrupole moment of the $d^{*}$ might be smaller or even change the sign. There is a calculation for the $d^*$ quadrupole deformation from the Beijing group~\cite{YB1}. In their case the main contribution to the $d^*$ quadrupole moment come from the interference between $S-wave$ $\Delta\Delta$ configuration and the small $D-wave$ $\Delta\Delta$ component of the $d^*$ wave function. In such an approach the $d^*$ should have a prolate shape with $Q_{d^*}= 0.025 efm^2$, similar in magnitude and sign to the result derived here from the $C$-term alone.

In the figure below the individual shapes of the $B$ and $C$-terms and the resulting distribution are plotted.

\begin{figure}[!h]
\begin{center}
\includegraphics[width=0.7\textwidth,angle=0]{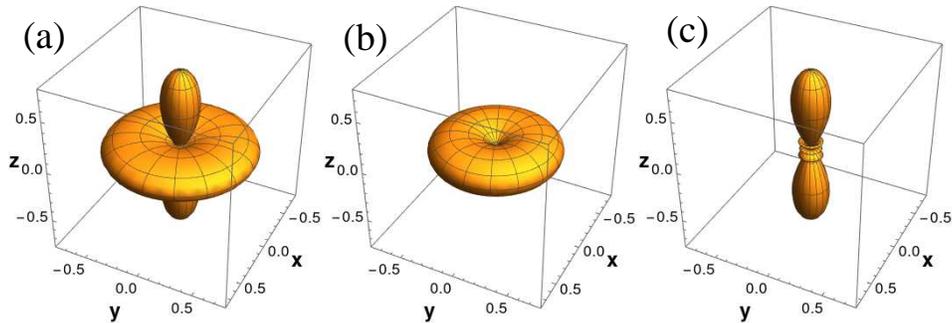}
\end{center}
\caption{$d^{*}(2380)$ quadrupole deformation produced by $N\Delta\pi$ ((b), $B$-term), middle; double-pionic cloud ((c), $C$-term), right and total effect, left, (a).}
\label{quadr}
\end{figure}

\section{Calculuation of the Transition quadrupole moment}

To calculate the $E2(\gamma d\rightarrow d^*)$ transition probability we need to calculate the $Q_{d\rightarrow d^*}$ quadrupole transition moment $Q_{d\rightarrow d^*}=<\Psi_d^*|\hat{Q}|\Psi_{d^*}>$. Before tackling the complicated $d\rightarrow d^*$ case, we first consider the deuteron itself. The wave function of the deuteron can be written as
\begin{eqnarray}
|\Psi_{d}> = \alpha|\Psi_{d}^S>+\beta|\Psi_{d}^D>, \alpha^2+\beta^2=1
\end{eqnarray}
here $|\Psi_{d}^S>$ is the $S-wave$ component of $pn$ inside the deuteron and $|\Psi_{d}^D>$ is the $D-wave$. The $D-wave$ probability in the deuteron is small, with $P_D=\beta^2\approx 4\%$. To calculate the deuteron quadrupole moment we need to fold the quadrupole operator into the deuteron wave functions $Q_d=<\Psi_d^*|\hat{Q}|\Psi_{d}>$. Since $\hat{Q}\sim Y^2_0$ then it follows that $<(\Psi_d^S)^*|\hat{Q}|\Psi_{d}^S>=0$ due to the sphericity of the S-wave component. The $<(\Psi_d^D)^*|\hat{Q}|\Psi_{d}^D>$ term would be suppressed by the low $P_D$ probability in the wavefunction, so the main contribution to the deuteron quadrupole moment arises from  the $<(\Psi_d^D)^*|\hat{Q}|\Psi_{d}^S>$ term. It should be noted that the nucleon has a small intrinsic quadrupole deformation~\cite{Buch1}, as the S-wave component is not fully spherical. However this is a small effect compared to the contributions calculated here and is neglected. 

The S-wave part of the $d^*$ is assumed to be essentially spherical, as also assumed in calculations of the deuteron.  Therefore the dominant term contributing to the transition quadrupole moment, and therefore the $E2$ transition probability,  is $<(\Psi_d^D)^*|\hat{Q}|\Psi_{d^*}>$. The $d^*$ is excited from the $D-wave$ part of the deuteron {\em only}. The deuteron to $d^*$ transition quadrupole moment can then be written as
\begin{eqnarray}
Q_{d\rightarrow d^*} = Q_d\frac{<R_{d^D}|R_{d^*}>}{<R_{d^D}|R_{d^S}>}
\end{eqnarray}
here $R_{d^S}$ and $R_{d^D}$ are the radial parts of a deuteron $S-wave$ and $D-wave$ wave function, and $R_{d^*}$ is a radial part of the $d^*$ wave function. By measuring the transition quadrupole moment we directly measure the $d^{*}$'s compactness in terms of the deuteron size. Therefore the strength of $d^*$ production via an $E2$ transition, which can be extracted in photoproduction experiments using double-polarisation measurements in which photon and deuteron spins are aligned,  have the potential to provide measurement of the $d^*$'s compactness. 
Taking the known deuteron wave function and combining it with $d^*(\Delta\Delta)$ wave function from Ref.~\cite{YB1} one can evaluate
\begin{eqnarray}
\frac{<R_{d^D}|R_{d^*}>}{<R_{d^D}|R_{d^S}>}\approx\frac{0.15}{0.22}\approx0.7~~and~~Q_{d\rightarrow d^*}\sim 0.20 efm^2
\end{eqnarray}
Note that neither the $S-wave$ or $D-wave$ part of deuteron can be excited directly into the $|6q>$ part of the $d^*$ (one cannot transfer two color bags into one with a colorless photon), so only the $\Delta\Delta$ part will be relevant to the $d^*$ production. If the $\Delta\Delta$ part is indeed $1/3$ of the $d^*$'s wave function, as predicted in Ref.~\cite{DBT9}, then the transition quadrupole moment would be further suppressed by this factor.

\section{Octupole magnetic moment}
As shown in Ref~\cite{Buch3} the octupole magnetic moment can be evaluated within a pion cloud model using the octupole moment operator 
\begin{eqnarray}
\hat{\Omega} = e_{\pi} \sqrt{\frac{16\pi}{5}}r_{\pi}^2 Y_0^2(\hat{r_\pi})\hat{\mu}\tau_z^N\sigma_z^N=\hat{Q}\hat{\mu}
\end{eqnarray}
Here the quadrupole operator acts on a pion cloud and the magnetic moment operator acts on the core. The quadrupole moment has been calculated in the previous section, so it is possible to calculate the magnetic moment of the $d^*$ in a simple quark model:
The $d^*$ wave function on a quark level can be written as 
\begin{eqnarray}
|d^*> = |u\uparrow u\uparrow u\uparrow d\uparrow d\uparrow d\uparrow>
\end{eqnarray}
Therefore the magnetic moment of the $d^*$ can be calculated as
\begin{eqnarray}
\mu_{d^*} = 3\mu_u+3\mu_d,~ \mu_u =\frac{2}{3}\frac{e\hbar}{2M_qc},~ \mu_d =-\frac{1}{3}\frac{e\hbar}{2M_qc} \nonumber \\
with~M_q=M_N/3,~\mu_u =2\mu_N,~\mu_d =-2\mu_N\nonumber \\
\mu_{d^*} = 3\mu_N\sim\mu_p 
\end{eqnarray}
To calculate the $d^*$ octupole moment we also need to evaluate the magnetic moment of a deuteron core and a $\Delta N$-core (the $C$ and $B$-terms in the quadrupole moment calculations respectively). The "deuteron core" has only a deuteron $S-wave$ component, so its magnetic moment would be $\mu_{C-term}=\mu_p+\mu_n=\mu_N$. For the $B$-term we have $\mu_{B-term}=\mu_{\Delta^0}+\mu_n=\mu_n=-2\mu_N$. The double-pion cloud  and the $N\Delta\pi$ terms have opposite signs. In quadrupole moment they cancel each other, but in octupole moment they add up. In ultimate situation we can have zero quadrupole moment and large octupole moment. Also the octupole moment of the $d^*$ is positive.
\begin{eqnarray}
\Omega_{d^*}=(Q_{d^*}^{C-term}-2Q_{d^*}^{B-term})\cdot \mu_N\approx -1.98Q_{\Delta^+}\mu_N=0.0089efm^3
\end{eqnarray}
In the work of Ref.~\cite{YB1} the octopole moment is calculated to be $\Omega_{d^*}=-0.00567efm^3$ - the opposite sign and smaller magnitude. However it should be noted that their magnetic moment is about twice as large $\mu_{d^*}=7.6\mu_N$, so having large magnetic moment and a $C$-term only one can reproduce the result of Ref.~\cite{YB1}. The estimation of magnetical octupole moment magnitude is important for the feasibility studies of upcoming experiments at photon beam facilities. While experimental determination of the $\Omega_{d^*}$ might be challenging with rather poor accuracy of the magnitude determination. The $\Omega_{d^*}$ sign evaluation should be straightforward. 

\section{Transition octupole magnetic moment}
The transition octupole moment can be calculated the same way as transition quadrupole moment. The arguments about the exclusive deuteron $D-wave$ contribution holds also for this case. The $D-wave$ part of of the deuteron wave function contributes to the magnetic moment in two ways - from the spin of the constituents and from the orbital motion of the charged proton $\mu_d^{D-wave}=-\frac{3}{2}(\mu_p+\mu_n-1/2)$. The orbital part is irrelevant for our calculations\footnote{it give rise to a structure $<Y_0^2|Y_0^1|Y_0^0>=0$} and the spin part has opposite sign due to anti-alignment of nucleon and deuteron spins in the case of the $D-wave$ component. Since we consider the deuteron to $d^*$ transition for the following spin state $|S_d=1,S^z_d=+1>\rightarrow |S_{d^*}=3,S^z_{d^*}=+1>$ we need to reevaluate the magnetic moment for another $d^*$ spin state
\begin{eqnarray}
|d^*>(S=3,S_z=1) =1/3(|u\downarrow u\downarrow u\uparrow d\uparrow d\uparrow d\uparrow>+|u\uparrow u\uparrow u\downarrow d\downarrow d\uparrow d\uparrow>+|u\uparrow u\uparrow u\uparrow d\uparrow d\downarrow d\downarrow>)
\end{eqnarray}
\begin{eqnarray}
\mu_{d^*}(S_z=1) =1/3( -\mu_u+3\mu_d+\mu_u+\mu_d+3\mu_u-\mu_d)=\mu_u+\mu_d =\mu_N 
\end{eqnarray}
The transition moment can be roughly evaluated as
\begin{eqnarray}
\Omega_{d\rightarrow d^*}=-Q_{d\rightarrow d^*}\sqrt{(\mu_p+\mu_n)\cdot\mu_d^*(S_z=1)}\sim -Q_{d\rightarrow d^*}\mu_N=-0.021 efm^3
\end{eqnarray}

Double-polarized photoproduction measurements with a tensor polarized deuteron target should be an ideal tool to access the $d^*$ magnetic octupole transition moment and the extraction of the $d^*$ magnetic moment. Such a target is planned to be used at CLAS(JLab) in near future. The experiments at CLAS are planned to be done with electron beam, allowing potential extraction of the magnetic octupole form-factor and not just $G_{M3}(0)$.

\section{The M3/E2 ratio}
It was demonstrated with $\Delta$ photoexcitation that the ratio of multipole transitions can be calculated and measured more precisely than the transitions themselves. For the case of $p\rightarrow\Delta$ transition the $E2/M1$ ratio can be accessed:
\begin{eqnarray}
\frac{E2}{M1}(p\rightarrow\Delta)=\frac{M_N\omega_\gamma Q_{p\rightarrow\Delta}}{6\mu_{p\rightarrow\Delta}}\approx 3\%
\end{eqnarray}
It is interesting to note that one of the first attempts to calculate $p\rightarrow\Delta$ $E2/M1$ ratio was done in the 60's using Weisskopf widths from nuclear physics gave remarkable agreement~\cite{E2Weiss}. 
\begin{eqnarray}
\frac{E2}{M1}(p\rightarrow\Delta)=\frac{2.4\cdot 10^{-8}R^4k^5}{2.1\cdot 10^{-2} k^3}=1.14\cdot10^{-6}R^4k^2\approx 3\%
\end{eqnarray}
here $R=0.8fm$ is the radius of proton and $k$ is the photon momentum. 
For the $M3/E2$ transition the Weisskopf width dependence would be~\cite{Krane}
\begin{eqnarray}
\frac{M3}{E2}(d\rightarrow d^*)\sim\frac{R^4k^7}{R^4k^5}= \Big (\frac{\omega_{\gamma}}{M_N} \Big)^2\cdot \Big (\mu_p-\frac{1}{4} \Big)^2\cdot \frac{8}{392}\approx 5\%
\end{eqnarray}
in which the size term cancels.
However from nuclear physics moments measurements we know that large deformation can lead to sizable deviation from Weisskopf coefficients. In such a case one needs to substitute them by a transition moments:

\begin{eqnarray}
\frac{B_w(M3)}{B_w(E2)} = \frac{250\cdot \mu_N^2}{144\cdot e^2}\approx 1.9 \cdot 10^{-2} \rightarrow 
\frac{\Omega_{d\rightarrow d^*}}{Q_{d\rightarrow d^*}}\approx 0.11
\end{eqnarray}

% \begin{eqnarray}
% \frac{M3}{E2}(d\rightarrow d^*) \sim 40\%
% \end{eqnarray}

% We observe similar behavior in our ansatz:
 \begin{eqnarray}
 \frac{M3}{E2}(d\rightarrow d^*) \sim \frac{\omega_{\gamma}^2\Omega_{d\rightarrow d^*}}{Q_{d\rightarrow d^*}} \sim \frac{\omega_{\gamma}^2Q_{d\rightarrow d^*}\mu_{d\rightarrow d^*}}{Q_{d\rightarrow d^*}}\sim \omega_{\gamma}^2\mu_{d\rightarrow d^*}\sim 30\% 
 \end{eqnarray}
In the $M3/E2$ ratio the size term from quadrupole moment cancels leaving a proportionality to the transition magnetic moment.
In the case of $d\rightarrow d^*$ one can asses one more type of excitation, $E4$, which may give senistivity to the $D-wave$ configuration inside the $d^*$. This could be accessed with double-polarized photoporduction measurements,  in which the deuteron and photon spins are anti-aligned. The production cross-section is expected to be small, but such information would be very valuable.
One word of caution need to be put here. In our calculations of transition quadrupole and octupole moments as well as  the $M3/E2$ ratio the main dependence come from deuteron. However if we compare the strength of transition moments with the $d^*$ moments we see a sizable difference:
\begin{eqnarray}
\Big \lvert \frac{\Omega_{d^*}}{\Omega_{d\rightarrow d^*}} \Big \rvert \approx 0.43, \Big \lvert \frac{Q_{d^*}}{Q_{d\rightarrow d^*}}\Big \rvert \approx 0.045
\end{eqnarray}

An order of magnitude difference in these ratios can point out us to a possible further suppression of $E2$ transition in favor of $M3$. If the $d^*$ magnetic moment is indeed $\mu_{d^*}=7.6\mu_N$ as anticipated in Ref.\cite{YB1}, the $M3/E2$ ratio gets as high as 80\%.

\section{Summary}

We have calculated the $d^*$ electromagnetic properties exploiting simple theoretical models. The quadrople and octopole moments were calculated in a pion cloud model and reasonable agreement was obtained with Resonating Group Method of Ref~\cite{YB1}.  The electromagnetic transition moments from the deuteron to the $d^*$ were also calculated for the first time. These results will help guide future experimental investigations of the $d^*$ with electromagnetic beams, where there is the potential to reveal important new information on the $d^*$ structure. Particularly sensitivities may be obtained from measurements of photoproduction from a tensor polarised deuteron target. We hope these initial theoretical studies will motivate more detailed theoretical work in the future.

\end{document}